# R-Index: A Metric for Assessing Researcher Contributions to Peer Review

Milad Malekzadeh[1]*

[1]Digital Geography Lab, Department of Geosciences and Geography,

University of Helsinki, Finland

*Corresponding author. Email: milad.malekzadeh@helsinki.fi;

https://orcid.org/0000-0003-2275-0497

**Abstract**

I propose the R-Index, defined as the difference between the sum of review responsibilities for a researcher's publications and the number of reviews they have completed, as a novel metric to effectively characterize a researcher's contribution to the peer review process. This index aims to balance the demands placed on the peer review system by a researcher's publication output with their engagement in reviewing others' work, providing a measure of whether they are giving back to the academic community commensurately with their own publication demands. The R-Index offers a straightforward and fair approach to encourage equitable participation in peer review, thereby supporting the sustainability and efficiency of the scholarly publishing process.

**Keywords**: Peer Review Contribution, R-Index, Academic Publishing Metrics, Researcher Assessment

## 1. Introduction

The peer review process is an integral part of academic publishing, ensuring the quality and integrity of scholarly work. However, the process is often slow, with researchers experiencing significant delays (Björk & Solomon, 2013). One major reason for these delays is the difficulty journal editors face in finding legitimate reviewers for submitted manuscripts (Dance, 2023). Researchers are typically burdened with their own research, teaching responsibilities, and administrative duties, leaving them with little time to engage in peer review. On the other hand, the demand for peer-reviewed publications continues to rise (Kovanis et al., 2016), driven by an academic culture that prioritizes publications in institutional evaluations (Bornmann & Daniel, 2006; Liu & Cheng, 2005). This pressure compels researchers to focus on publishing more frequently, further exacerbating the strain on the peer review system.

Despite numerous critiques of the current publishing process (Brembs et al., 2023), the value of peer review remains undeniable. Peer review provides an essential check on the quality and validity of research, offering a second set of eyes to evaluate the rigor and significance of the work. Yet, a significant issue remains: many researchers are prolific in their publication output but seldom contribute to the peer review process. Existing metrics in academia, like the h-index (Hirsch, 2005), measure the impact of a researcher's work through citations. While platforms like Publons (Teixeira da Silva & Nazarovets, 2022) assess researchers' contributions to the peer review process, they do so in a one-dimensional manner. These metrics disregard the number of papers a researcher has published and the subsequent burden these papers impose on the review system. This oversight leaves a significant gap in the current scientific contribution metrics, failing to fairly measure the responsibility researchers have to contribute to the peer review process. In this paper, I propose the R-Index, a novel metric designed to quantify and balance the contributions of researchers to the peer review system relative to their publication output.

## 2. The Need for a Balanced Contribution Metric

A single paper may typically undergo two rounds of review, each involving two to three reviewers. On average, each scientific publication requires 3.49 ± 1.45 (SD) reviews (Raoult, 2020). Publishing multiple papers per year, each requiring multiple rounds of review, can impose a significant burden on the peer review system. This cumulative demand on reviewers can lead to substantial delays and inefficiencies within the academic publishing process. For example, if a researcher publishes five papers in a year, this could equate to approximately 17 reviews on average (with a range of about 10 to 25 reviews) being conducted for their work alone. The current system lacks a standardized measure to determine whether researchers are giving back to the community by participating in peer review. Simply counting the number of reviews completed by a researcher is inadequate, as it fails to account for the volume of review work their publications demand from the community. A more rigorous approach is needed to assess whether researchers are contributing their fair share to the peer review process.

## 3. Conceptualizing the R-Index

The R-Index is designed to measure a researcher's contribution to the peer review process relative to the demands their publications place on the system. The index aims to balance the number of reviews a researcher completes with the number of reviews their publications necessitate. The calculation of the R-Index involves several steps to ensure a fair and accurate representation of a researcher's contributions.

1. Calculating Review Responsibility: For each paper, determine the total number of reviews it received. Then, divide this total by the number of authors to calculate each author's review responsibility. For example, if a paper received six reviews and there are four authors, the review responsibility per author is 1.5 reviews (6 reviews / 4 authors).
2. Summing Review Responsibilities: Sum the review responsibilities for all papers a researcher has published within a given period to determine the total expected reviews.
3. Summing Review Contributions: Sum the number of reviews the researcher has completed in the same period
4. Calculating the R-Index: The R-Index is calculated by subtracting the total number of reviews completed by the researcher from the total review responsibility derived from their publications. The formula for the R-Index is:

$$R - Index = Total\ Reviews\ Completed - \sum_{p}(\frac{Total\ Reviews\ Received}{Number\ of\ Authors})_p$$

where $p$ represents each published paper. A positive R-Index indicates that the researcher has contributed more to peer review than their publications have received, while a negative R-Index indicates that the researcher has not contributed enough. An R-Index of zero suggests an equal balance between contributions and demands.

## 4. Potential Benefits

The introduction of the R-Index has several potential benefits for the academic community. First, it provides a tangible measure of a researcher's contribution to the peer review process, encouraging a more balanced participation. Researchers with low R-Indices might be motivated to increase their reviewing activities, knowing their contributions are being quantified and recognized.

Second, the R-Index could help alleviate the burden on journal editors who struggle to find reviewers. By promoting a culture of reciprocal reviewing, the overall efficiency of the peer review process could improve, reducing delays in manuscript evaluations and accelerating the dissemination of scientific knowledge.

Third, the R-Index offers a transparent and fair way to assess reviewing contributions, independent of factors such as field of study or research impact, which influence other metrics like the h-index. It is a metric that researchers can directly control by choosing to engage more actively in the peer review process.

## 5. Implementation and Challenges

Journals can calculate and report the review responsibility for each author by dividing the total number of reviews received by the number of authors. This information, combined with data from platforms such as Publons, allows for the calculation of the R-Index for each researcher. The R-Index can then be reported alongside other metrics, such as the h-index, to provide a more comprehensive assessment of a researcher's contributions to academia.

Despite its potential benefits, the implementation of the R-Index presents several challenges. Coordinating the calculation and reporting of review responsibilities across various journals and disciplines requires significant collaboration. Additionally, integrating data from multiple platforms like Publons into a consistent and reliable framework may be logistically complex.

Early career researchers may not be frequently invited to review articles, particularly when they are just beginning their academic careers. Editors typically invite reviewers with an established publication record or based on referrals, which can limit the opportunities for those new to the field to participate in the peer review process. This lack of engagement could negatively impact their reputation if their R-Index is calculated immediately following their first publications. To mitigate this issue, I propose implementing a two-year lag in the calculation of the R-Index. This adjustment period would provide researchers with sufficient time to engage in the review process, allowing them to contribute to the system in proportion to their publication output. By incorporating this lag, the R-Index can more accurately capture a researcher's long-term commitment to peer review, ensuring that early career researchers are not unfairly penalized during the initial stages of their academic careers.

The quality of peer reviews can differ greatly; while many reviewers dedicate substantial time and effort to providing thorough assessments, a minority may offer only minimal or superficial feedback. To prevent the misuse of the R-Index by inflating review counts with low-quality contributions, a mechanism could be implemented where editors have the discretion to exclude subpar reviews from being counted toward a researcher's R-Index. This would ensure that the metric reflects meaningful and constructive contributions to the peer review process.

Reviewing academic papers is not the only form of contribution to the review process that researchers provide to the scholarly community. Senior researchers are often invited to review grant and research proposals for various institutions, a task that can be significantly more time-consuming than reviewing manuscripts. However, this process differs from peer review of academic papers, as those reviewing grant proposals are often also eligible to apply for grants themselves, potentially balancing their contribution to the system. Although these contributions are valuable, they are difficult to quantify and may vary depending on the researcher's career stage. A multifaceted R-Index could account for different types of review contributions by assigning distinct values to various review activities. However, developing such a complex metric extends beyond the scope of this initial conceptualization of the R-Index.

Another significant contribution, particularly by senior researchers, is serving as editors of academic journals. This role can be time-consuming and may limit their ability to conduct peer reviews. However, many journals report the editorial contributions and reviews performed by editors. Whether each round of revisions is counted as a separate review or each paper is counted as a single review, these contributions can be reflected in the researcher's R-Index. Addressing these nuances is crucial to ensuring the R-Index accurately and effectively evaluates researchers' overall contributions to the peer review process.

A final consideration is that the R-Index is not designed to impose additional time burdens on researchers but rather to promote a more balanced and equitable peer review system. The metric encourages researchers to align their publication output with their contributions to the review process. This balancing act could serve as a moderating influence on those who are inclined to publish prolifically, particularly in environments where institutional evaluations prioritize the quantity of publications over their quality. By integrating the R-Index into the academic landscape, researchers would be aware that they need to allocate sufficient time to peer review activities, ensuring that their engagement with the system is proportionate to the demands their publications place on it. This approach does not suggest that researchers need to increase their overall workload, but rather that they should balance their use of the system by contributing appropriately to its maintenance and sustainability.

## 6. Conclusion

The R-Index represents a novel approach to balancing the demands and contributions within the peer review system. By quantifying the difference between the number of reviews completed and the review responsibility derived from publications, it provides a clear and fair measure of a researcher's engagement in this crucial aspect of academic publishing. Adoption of the R-Index could foster a more equitable and efficient peer review process, benefiting researchers, editors, and the broader scientific community.

In conclusion, the R-Index addresses a critical gap in the current academic landscape by recognizing and incentivizing the essential contributions of peer reviewers. It aligns the responsibilities of researchers with their participation in the peer review process, ensuring that the system remains sustainable and effective for future generations of scholars.

**Funding statement**

This study has not received any funding.

**Conflict of interest disclosure**

No potential conflict of interest was reported by the author.

**Ethics approval statement**

This study did not require any ethics approval.